\documentclass[12pt]{article}
\usepackage{verbatim,graphicx,amssymb,cite}
\newcommand{\be}{\begin{equation}}
\newcommand{\ee}{\end{equation}}

\begin{document}
\setcounter{page}{1}
\begin{center}
{\Large \bf Possible trace of neutrino nonstandard interactions in the supernova
}\\
\vspace{0.5cm}

{\large
C. R. Das\footnote{E-mail: crdas@cftp.ist.utl.pt},
Jo\~{a}o Pulido\footnote{E-mail: pulido@cftp.ist.utl.pt}\\
\vspace{0.15cm}
{{\small \sl CENTRO DE F\'{I}SICA TE\'{O}RICA DE PART\'{I}CULAS (CFTP)\\
 Departamento de F\'\i sica, Instituto Superior T\'ecnico\\
Av. Rovisco Pais, P-1049-001 Lisboa, Portugal}\\
}}
\vspace{0.25cm}
\end{center}

\begin{abstract}

Neutrino non-standard interactions (NSI), previously introduced for the sun, are studied
in the supernova context. For normal hierarchy the
probability for electron neutrinos and antineutrinos at low energy ($E\lesssim
0.8-0.9 MeV$) is substantially
increased with respect to the non-NSI case and joins its value for inverse
hierarchy which is constant
with energy. Also for inverse hierarchy the NSI and non-NSI probabilities are
the same for each neutrino and
antineutrino species. These are the possible visible effects of NSI in the
supernova.
The decay into antineutrinos, which has been previously shown to be
implied by dense matter, cannot be seen experimentally, owing to the smallness of the 
antineutrino production probability. 
\end{abstract}

\begin{figure} [htb]
\centering
\includegraphics[height=80mm,keepaspectratio=true,angle=0]{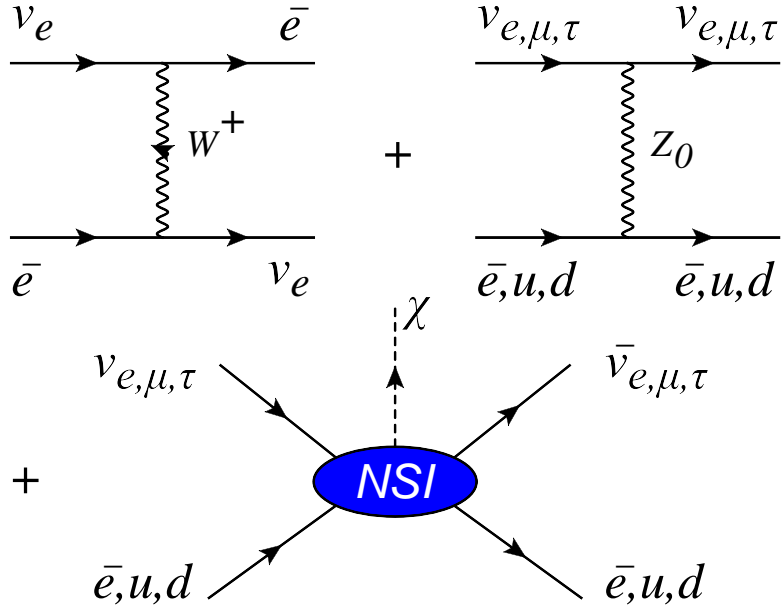}
\caption{\it The processes involved in the propagation of neutrinos in the sun as
in our model: the two upper diagrams are the standard ones for matter oscillation and
the lower one represents the decay $\nu_i\rightarrow \bar\nu_{j}+majoron (\chi)$.}
\label{fig1}
\end{figure}

It was shown some time ago \cite{Pulido:2010ht} that neutrino nonstandard interactions (NSI) 
provide a solution to the discrepancy between the observed flatness of the Super-Kamiokande 
electron spectrum \cite{:2008zn} and the large mixing angle (LMA) prediction while keeping all 
others \cite{Aharmim:2009gd,Bellini:2010gn,Cleveland:1998nv} accurate and in particular improving 
the Chlorine one \cite{Cleveland:1998nv}. As neutrinos propagate through matter they interact 
with electrons and quarks and their standard interaction potential is well known
\begin{equation}
V(SI)=V_c+V_n=G_F\sqrt{2}N_e\left(1-\frac{N_n}{2N_e}\right)
\label{VSI}
\end{equation}
with $V_c=(V_e)_{CC}=G_F\sqrt{2}N_e$ (CC contribution from electrons),
$V_n=-(G_F/\sqrt{2})N_n$ (contribution from neutrons, NC only). Our approach to NSI
assumes extra contributions to the vertices $\nu_{\alpha} \nu_{\beta}$ and $\nu_{\alpha}e$.
Denoting by $\varepsilon^{e,u,d}_{\alpha,\beta}$ the NSI factor that multiplies each diagram
associated to neutrino propagation in matter we have
\begin{eqnarray}
(v_{\alpha\beta})_{NSI}\!\!\! & = & \!\!\!G_F\sqrt{2}N_e\left[(\varepsilon_{\alpha \beta}^{eP})_{NC}+\left(-\frac{1}{2}+2sin^2\theta_W \right)
(\varepsilon_{\alpha \beta}^{eP})_{NC}+\left(1-\frac{8}{3}sin^2\theta_W+\frac{N_n}{2N_e} \right) 
\varepsilon_{\alpha \beta}^{uP}\right.\nonumber\\
&& + \left.\left(-\frac{1}{2}+\frac{2}{3}sin^2\theta_W-\frac{N_n}{N_e} \right)
\varepsilon_{\alpha \beta}^{dP} \right]
\label{VNSI}
\end{eqnarray}
where ${\alpha,\beta}$ are neutrino flavour indeces. The full matter potential is
therefore
\begin{equation}
V=V(SI)+V(NSI)
\label{V}
\end{equation}
and the matter Hamiltonian in the flavour basis is
\begin{equation}
{\cal H}_M=V_c \left(\begin{array}{ccc} 1 & 0 & 0\\
0 & 0 & 0\\
0 & 0 & 0\\ \end{array}\right)+
\left(\begin{array}{ccc} (v_{ee})_{NSI} & (v_{e\mu})_{NSI} & (v_{e\tau})_{NSI}\\
(v_{\mu e})_{NSI} & (v_{\mu\mu})_{NSI} & (v_{\mu\tau})_{NSI}\\
(v_{\tau e})_{NSI} & (v_{\tau\mu})_{NSI} & (v_{\tau\tau})_{NSI}\\ \end{array}\right)={\cal H_{SI}}+{\cal H_{NSI}}~.
\label{HNSI}
\end{equation}
In ref.\cite{Pulido:2010ht} it was found that only imaginary diagonal couplings $\varepsilon$ provide 
the change in the LMA survival probability in order to make it flat. To this end the necessary and sufficient
order of magnitude is found to be 
\begin{equation}
|\varepsilon_{ee}^{e,u,d}|\simeq|\varepsilon_{\mu\mu}^{e,u,d}|\simeq|\varepsilon_{\tau
\tau}^{e,u,d}|=(2-4)\times 10^{-4}.
\end{equation}
This indicates that the eigenstates described by the Hamiltonian (\ref{HNSI}) are unstable, this
instability obviously being induced by dense matter and interpreted as associated to the neutrino decay
into an antineutrino and a majoron\cite{Pulido:2010ht}. In mass eigenstates
\begin{equation}
\nu_{i}\rightarrow \bar\nu_{j}+\chi.
\end{equation} 
The processes involved in the propagation of neutrinos in the Sun are shown in fig.\ref{fig1}.

\begin{figure}\centering
\includegraphics[height=110mm,keepaspectratio=true,angle=-90]{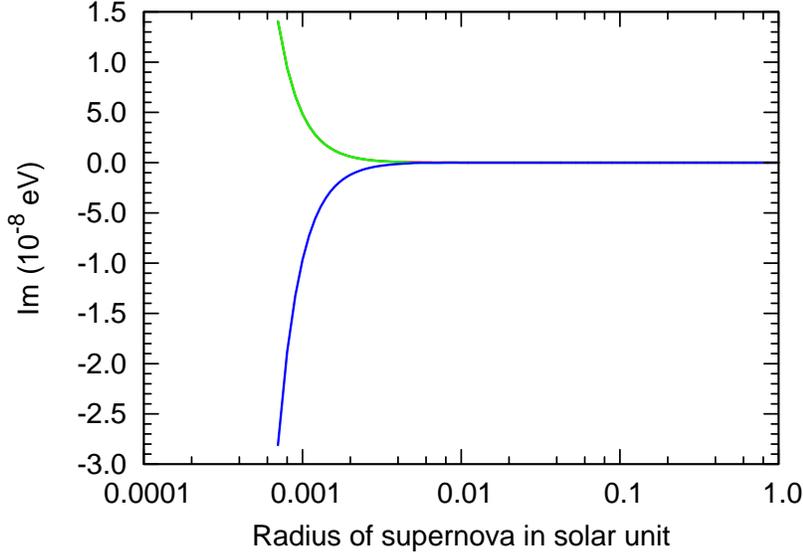}
\caption{\it The decay rates of the neutrino mass eigenstates.}
\label{fig2}
\end{figure}

The imaginary parts of the complex Hamiltonian eigenvalues give the decay rates of the neutrino mass
eigenstates which are quite small: the average lifetime in the initial $10^{-3}$ solar radius fraction 
is approximately $10^{-7}s$ and rises very rapidly afterwards (see fig.\ref{fig2}).

\begin{figure}[ht]
\centering
\vspace*{-5.0cm}
\hspace*{-1.3cm}
\includegraphics[height=180mm,keepaspectratio=true,angle=270]{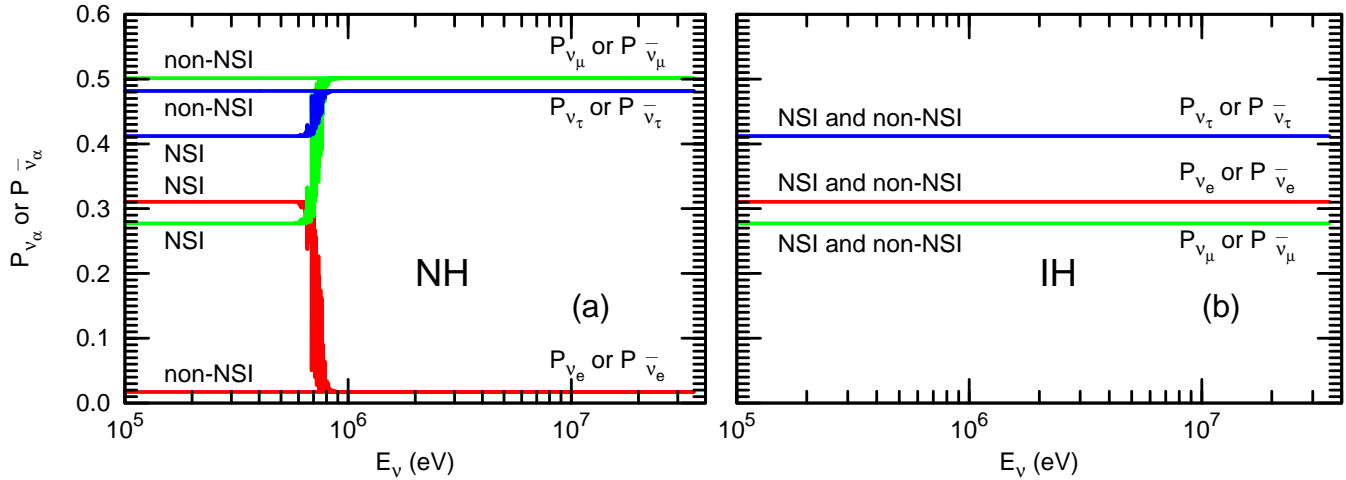}
\caption{ \it The appearance probabilities $P_{\nu_e},~P_{\bar\nu_e},~P_{\nu_{\mu}},
~P_{\bar\nu_{\mu}},~P_{\nu_{\tau}},~P_{\bar\nu_{\tau}}$
with and without NSI ((a) normal hierarchy, (b) inverse hierarchy). For normal hierarchy
and energy below 0.9 MeV the NSI probability merges with its non-NSI value.}
\label{fig3}
\end{figure}

The decay rate for the process $\nu_{i}\rightarrow \bar\nu_{j}+\chi$ satisfies
\begin{equation}
\frac{\partial \Gamma_{i}}{\partial E_f}=\sum_{j=1}^{3}\frac{|g_{ij}|^2}{8\pi}\frac{E_0-E_f}{{E_0}^2}
|v_{i}(r)-\overline v_{j}(r)|_{NSI}F(r,E_0)
\end{equation}
where $E_0,E_f$ are the initial neutrino and final antineutrino energies and $v_{i}=-\bar v_{i}$ are the interaction 
potentials in the mass basis. However for the neutrino-majoron couplings only upper bounds exist \cite{Farzan:2002wx}
\begin{equation}
\sum_{\alpha}|g_{e\alpha}|^2<5.5\times 10^{-6},~ \sum_{\alpha}|g_{\tau\alpha}|^2<5.5\times 10^{-2}
\end{equation}
The appearance probabilities for $\nu_{\alpha},\bar\nu_{\alpha}$ with and without NSI are shown in fig.\ref{fig3} 
for normal (panel (a)) and inverse hierarchy (panel (b)). For $\nu_e$ and $\bar\nu_e$ which have charged currents
and are therefore easier to detect, these probabilities are rather small in the absence of NSI and in normal 
hierarchy (fig.\ref{fig3}a). Their actual detection will crucially depend on the detector size and supernova distance. 
With NSI, they may become clearer through $\nu_e d\rightarrow ppe^{-}$, $\bar\nu_e p\rightarrow n e^{+}$ or an increased 
$\nu e^-\rightarrow \nu e^-$ or $\bar\nu e^-\rightarrow \bar\nu e^-$ scattering event rate, however at low energy
($E_0\lesssim 0.8-0.9MeV$), which is still an experimental challenge. For inverse hierarchy, as also pointed above,
the situation is much different: the $\nu_e$, $\bar\nu_e$ signal appears louder and clearer (fig.\ref{fig3}b).  

Finally the possibility of detection and measurement of the $\nu_{\mu}$, $\bar\nu_{\mu}$, $\nu_{\tau}$, 
$\bar\nu_{\tau}$ energy spectra, which can only be traced via neutral currents, may in future be made through 
$\nu~p\rightarrow\nu~p$ scattering in scintillator detectors (e.g. Borexino, SNO+) as proposed in 
\cite{Beacom:2002hs} and recently revived in \cite{Dasgupta:2011wg}. For the 
NSI scenario expound in the present paper this technique will be particularly useful, since it appears to be 
possible to clearly distinguish between normal and inverse hierarchies (see fig.\ref{fig3}). In fact it suffices 
to note that for normal hierarchy the above mentioned neutrinos arrive copiously on Earth in comparison with 
the more rare $\nu_e$$'$s and $\bar\nu_e$$'$s whereas for inverse hierarchy all species arrive in comparable 
numbers.


\section*{Acknowledgements}
CRD acknowledges a scholarship from the portuguese FCT (ref. SFRH/BPD/41091/2007). The authors are partially
supported by FCT through CERN/FP/116328/2010 and CFTP-FCT Unit 777 which are partially funded through POCTI (FEDER).


\end{document}